Invited Paper

## Title

Programming the scalable optical learning operator with spatial-spectral optimization

## Authors

Yi Zhou,[1,2,3†] Jih-Liang Hsieh,[1,2*†] Ilker Oguz,[1,2] Mustafa Yildirim,[1,2] Niyazi Ulas Dinc,[1,2] Carlo Gigli,[2] Kenneth K. Y. Wong,[3] Christophe Moser,[1] and Demetri Psaltis[2]

## Affiliations

[1] Laboratory of Applied Photonics Devices, École Polytechnique Fédérale de Lausanne (EPFL), CH-1015 Lausanne, Switzerland.
[2] Optics Laboratory, École Polytechnique Fédérale de Lausanne (EPFL), CH-1015 Lausanne, Switzerland.
[3] Department of Electrical and Electronic Engineering, The University of Hong Kong, Pokfulam Road, Hong Kong.
*Corresponding author. E-mail: jih-liang.hsieh@epfl.ch
†These authors contributed equally to this work.

## Abstract

Electronic computers have evolved drastically over the past years with an ever-growing demand for improved performance. However, the transfer of information from memory and high energy consumption have emerged as issues that require solutions. Optical techniques are considered promising solutions to these problems with higher speed than their electronic counterparts and with reduced energy consumption. Here, we use the optical reservoir computing framework we have previously described (Scalable Optical Learning Operator or SOLO) to program the spatial-spectral output of the light after nonlinear propagation in a multimode fiber. The novelty in the current paper is that the system is programmed through an output sampling scheme, similar to that used in hyperspectral imaging in astronomy. Linear and nonlinear computations are performed by light in the multimode fiber and the high dimensional spatial-spectral information at the fiber output is optically programmed before it reaches the camera. We then used a digital computer to classify the programmed output of the multi-mode fiber using a simple, single layer network. When combining front-end programming and the proposed spatial-spectral programming, we were able to achieve 89.9% classification accuracy on the dataset consisting of chest X-ray images from COVID-19 patients. At the same time, we obtained a decrease of 99% in the number of tunable parameters compared to an equivalently performing digital neural network. These results show that the performance of programmed SOLO is comparable with cutting-edge electronic computing platforms, albeit with a much-reduced number of electronic operations.

**Introduction**
Growing demand for artificial intelligence (AI) has evolved dramatically in recent years. The AI revolution is fueled by the immense parallel computing power of electronic hardware such as field-programmable gate arrays and graphic and tensor processing units [1–4]. However, the performance is inherently limited by the fundamental tradeoff between energy efficiency and computing power in electronic computing [5]. The "bigger is better" mentality has dominated, but the energy required to operate large networks becomes a limiting factor. Moreover, as the scale of an electronic transistor approaches its physical limit, it is necessary to investigate and develop new computing processors during the post-Moore's law era [6,7]. Optical computing, using photons instead of electrons as the information carrier, has the potential to provide high power efficiency, and high-speed processing [8,9].

The development of optical computing platforms has recently been the focus of intense research and commercial interest. Various optical neural network (ONN) architectures have been proposed, including diffractive optical neural networks [10,11], photonic reservoir computing [12,13], photonic spiking neural networks [14,15], and optical convolutional and recurrent neural networks [16,17]. An ONN with a computational speed of more than 10 TOPS has been demonstrated with today's advanced optical technologies [18]. The low energy consumption of ONNs is the key advantage of the technology. In fact, less than one photon per operation has been demonstrated [19], which can lead to orders of magnitude lower energy consumption than digital computation. Adaptive training of the ONNs with an error backpropagation algorithm is the most common method to guarantee the model's reliable and accurate network inference. Examples of trained ONNs include holographic ONNs [20], physics-aware training of an optoelectronic network [21], a reconfigurable diffractive optoelectronic neural network with three layers [22], and hybrid training of optical neural networks [23].

Scalable Optical Learning Operator (SOLO) was previously proposed by our group to utilize the optical nonlinear propagation of a graded-index multimode fiber (GRIN MMF) to transform the input dataset into a less demanding representation space [24]. This architecture is inspired by the Extreme Learning Machine algorithm [25], in which a fixed nonlinear mapping of the input data is combined with a simple single trainable layer. In SOLO, the data is optically encoded via a spatial light modulator (SLM) and the nonlinear propagation inside an MMF realizes the nonlinear mapping. The advantage of performing the nonlinear mapping optically is that it can reduce the number of trainable parameters in a neural network and provides reasonable performance on many different datasets. The nonlinearity of the MMF affects the output beam shape and broadens the spectrum. As the information bandwidth increases, valuable information may be spatially or spectrally hidden in the output beam. The SOLO system collects the output beam simply with a camera, for which the number of pixels is much smaller than the spatial-spectral information contained in the beam, thus omitting the potential to further increase the classification performance. It can be expected that the system's classification accuracy depends strongly on the measuring approach. Therefore, we expect that optimizing the sampling of the MMF output beam could improve the performance of the system.

In this paper, we implemented the back-end programmed-SOLO (P-SOLO) system and tackled the image classification task on COVID-19 dataset. A digital micromirror device (DMD) placed at the output of the MMF is used to select spatial features by turning DMD pixels on or off. A chromatic dispersion grating is placed after the DMD to disperse the sampled light to achieve an optimized spatial-temporal measurement by the camera. We ran two sampling approaches. In the first approach (multi-line), we sampled the output beam of the MMF as a line, which is perpendicular to the orientation of the dispersion grating vector. The sampling line is swept across the DMD and capturing the sampled beam on the camera simultaneously. This generates multiple sampled images per input. In the second approach (single-shot), we optimized the DMD sampling pixels that generated only one sampled image per input. In the final experiment, we combined the front-end programming described in [26] with the proposed single-shot approach, and we were able to generate the best classification performance (89.9%) of the COVID-19 dataset on an optical network.

## Materials and Methods

**Experimental setup:** The P-SOLO system includes an ultrafast laser source (Amplitude Laser Satsuma), whose pulses could be adjusted to have pulse widths from 700 fs to 10 ps and 125 kHz repetition rate. The pulse has a center wavelength of 1034 nm and spectral width of 7 nm. The linearly polarized laser output beam with a Gaussian beam shape encodes the dataset via a phase-only Spatial Light Modulator (SLM, Meadowlark). The SLM is a reflective liquid crystal on silicon, with 1920×1152 pixels, placed with a pitch of 9.2 μm and an 8-bit dynamic range operated at a frame rate of 50 Hz. The laser beam was expanded to cover 520×520 pixels on the SLM, which was used to input the dataset images. The pulses then propagate in a 5 m GRIN 50/125 MMF with a numerical aperture (NA) of 0.2 that supports 240 modes at the operation wavelength. The phase-modulated light from the SLM is coupled to the MMF through a lens with a focal length of 36 mm. The coupled beam covers the whole MMF core area. The output beam pattern from the MMF was collimated and then directed to a DMD. The DMD (Ajile AJD-4500) consists of 912×1140 micromirrors with a pitch of 7.6μm. The light sampled and reflected by the DMD is then dispersed by a diffraction grating with a 1200line mm$^{-1}$ period. The dispersed beam after the grating is recorded by a camera with a 6.9 μm pixel pitch and a maximum frame rate of 552 Hz (BFS-U3-04S2M-CS). We program the DMD mask through a surrogate optimization algorithm [27] to find the optimal spatial sampling pattern. The beam is attenuated with an adjustable neutral density filter before camera detection. The input pulse peak power was optimized to maximize the nonlinear interaction of multimodes in MMF as in the SOLO system [24].

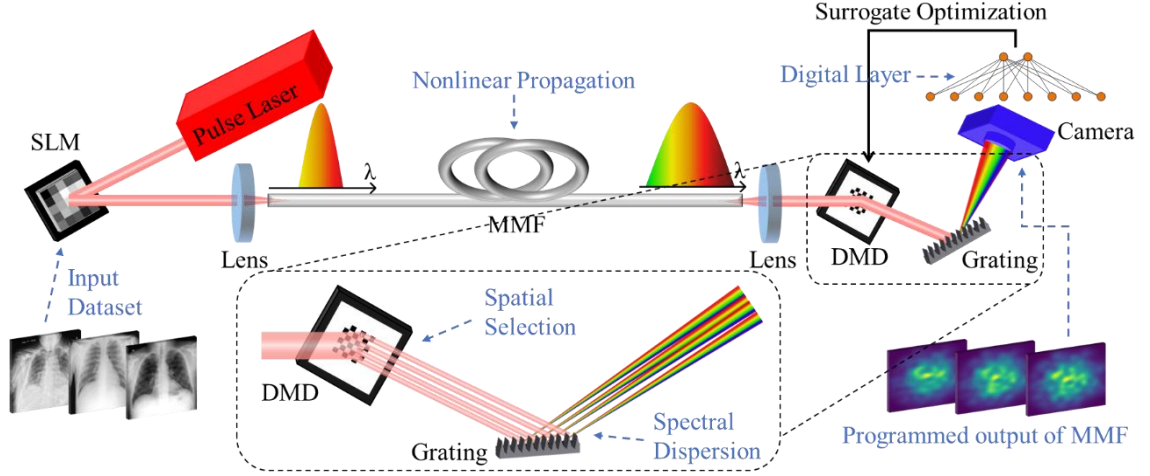

**Fig. 1.** Schematic of P-SOLO experimental setup. The experimental setup comprised an SLM for encoding dataset images onto laser pulses, an MMF for nonlinear transformation with spatial-temporal optical nonlinearities, a DMD as a space mask for programming combined with a surrogate optimization algorithm to extract spatial feature map, and a grating as the spectral frequency-resolved beam profile measurement technique to further improve classification accuracy.

**Multi-line mask sampling approach**

We first performed a sequential line sweep method across the output beam after the MMF. The COVID-19 dataset [28] was served as comparing metrics with 3000 X-ray samples (2400 X-ray samples for training and 600 X-ray samples for testing). The accuracy previously reported in SOLO was 83% [24]. The X-ray sample is encoded on the phase of the input beam through the SLM while the corresponding fiber output intensity patterns are spatially sampled by the DMD. The distance of the DMD, grating, and camera are 54 cm and 17 cm respectively (Fig. 2a). The line width of DMD is a parameter that needs to be optimized. In the first experiment, the line mask of DMD is designed to be perpendicular to the dispersion grating vector, meaning that the spatial sampling and spectral sampling are perpendicular to each other. The vertical DMD line masks, found to be optimized at 48 pixel width (see Supplementary), are horizontally swept at 4 sequential location to spatially sample the MMF output (4H configuration, Fig. 2b). One sample in the dataset will generate 4 intensity images recorded by the camera. After concatenating these 4 images, the digital layer obtains a test accuracy of 86.2% for 4H configuration.

In the second experiment, the DMD line masks are rotated. The horizontal DMD line masks, found to be optimized at 58 pixel width (see Supplementary), are vertically swept at 5 sequential locations and the camera recorded 5 corresponding intensity image (5V configuration, Fig. 2c). The test accuracy for 5V configuration is 85.9%. In 5V configuration, spatial sampling and spectral sampling are parallel to each other, thus feature information hidden in the beam are not extracted as well as 4H configuration. However, with the help of free-space diffraction and grating dispersion, this accuracy is still higher than the 83% reported previously from the SOLO system [24]. If we further combine the 9 images from 4H and 5V configuration (4H+5V configuration, Fig. 2d), the classification accuracy of P-SOLO can go up to 88%, surpassed the 87.3% accuracy obtained by LeNet-5 (82826 training parameter) [29] digital neural networks. The improvement is summarized in Fig. 3 together with the accuracy previously reported in SOLO [24].

The distance between the DMD, grating, and camera, as indicated in Fig. 2a plays an important role since it determines the degree to which the nonlinear spectral components of the field propagate in free-space and are separately recorded on the camera. In the third experiment, we manually optimized the distances between optics components, and we were able to improve the classification accuracy for 4H configuration from 86.2% to 88% (4H+optimized distance configuration, Fig. 4). This accuracy is as high as the 4H+5V configuration while it only required concatenating 4 images instead of 9 images. This could reduce the number of training parameters in the digital decision layer of P-SOLO and shorten the experiment time. This result also demonstrates that the free-space diffraction also plays a vital role in the spatial-spectral sampling in the P-SOLO system.

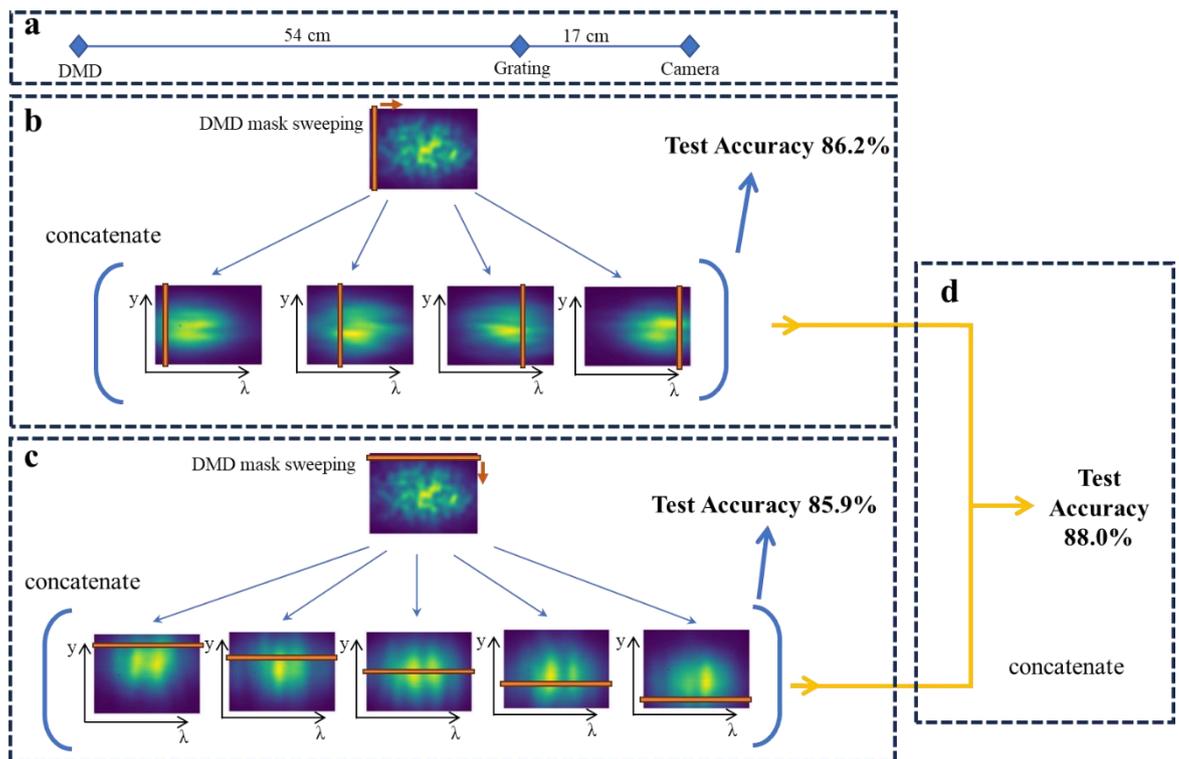

**Fig. 2.** Multi-line mask sampling approach with different experimental configurations. (a) distance between the DMD, the grating and the camera. (b) 4 horizontally sweeping vertical DMD lines (4H configuration). (c) 5 vertically sweeping horizontal DMD lines (5V configuration). (d) Combing both DMD line orientation (4H+5V configuration).

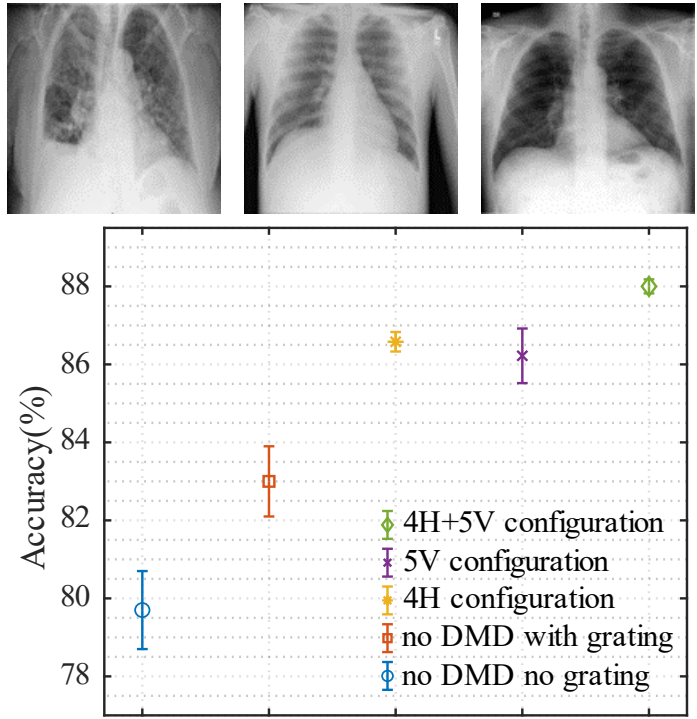

**Fig. 3.** Comparison of classification accuracy for the COVID-19 X-ray dataset under different multi-line mask sampling. 4H, 5V, and 4H+5V are the method proposed in this study. No DMD no grating and no DMD with grating are previously reported in SOLO [24].

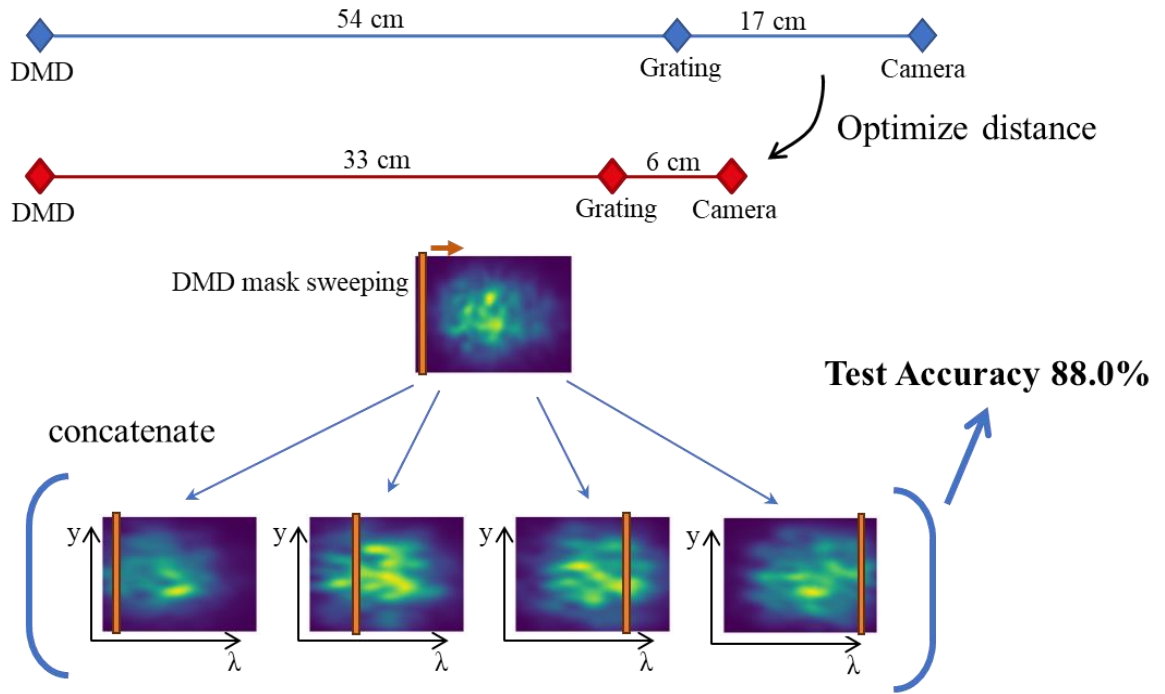

**Fig. 4.** The classification accuracy improvement after optimizing the distance between the DMD, the grating, and the camera. The optimized distance is shown on the top of the figure. The accuracy obtained is as high as 4H+5V configuration while only required 4 images instead of 9 images.

**Single-shot mask approach**

Next, we optimized the sampling operation by programming a single DMD mask. The DMD is divided into several superpixels and the MMF output beam covers 80 superpixels. To avoid overlapping between the superpixels at the camera plane, a spacing parameter between the superpixel (zero-padding) was first introduced. A smaller COVID-19 dataset was used (1200 X-ray samples for training and 300 X-ray samples for testing) to efficiently determine the spacing parameter of DMD superpixel. A total of 80 (4×20) superpixels are encoded on the DMD, each superpixel consisting of multiple DMD binary pixels. The height of the superpixel contains 30 DMD pixels. The width of the superpixel is a parameter that needs to be optimized in a range from 16 to 48 DMD pixels and it is determined through surrogate optimization. Specifically, 80 parameters in the optimization algorithm control the on/off of the DMD superpixels, and another 5 binary parameters control the width of the superpixel (a variable between 16 to 48). During surrogate optimization, the classification accuracy is generated by the digital decision layer and acts as the objective function to be optimized.

The result of the optimized mask indicates that the best width for each superpixel is 48 pixels, which means there is no zero-padding between each superpixels. We next applied optimized DMD mask on full size COVID-19 dataset (3000 X-ray samples of COVID-19 dataset with 2400 training samples and 600 testing samples). The iteration of the surrogate optimization is shown in Fig. 5. The accuracy gradually increases with the number of iterations and the best test accuracy is found to be 86.9% after 400 iterations. The scatter in Fig.5 indicates the testing accuracy as a function of each iteration number and the blue dashed line represents the best accuracy obtained up to this point. The initial and optimized DMD masks (black superpixel corresponds to the on state and white superpixel corresponds to the off state) are also shown in Fig. 5.

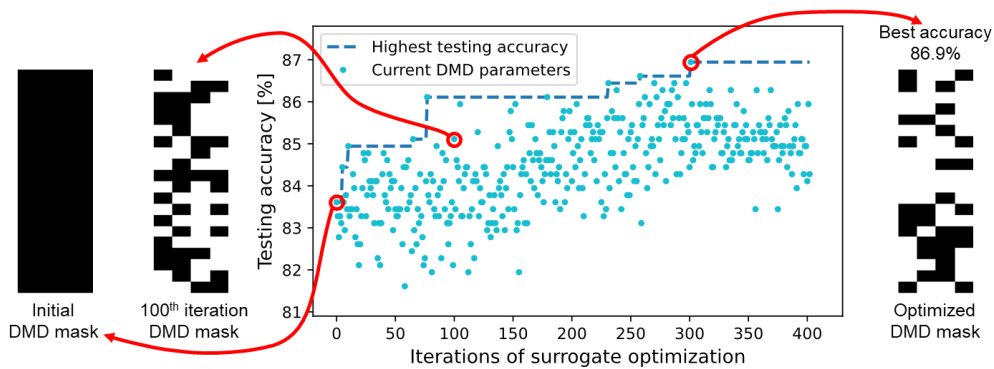

**Fig. 5.** Classification accuracy evolution of 3000 samples of COVID-19 X-ray dataset by programming single-shot DMD mask with surrogate optimization. Single-shot DMD mask divides into four lines including 80 superpixels with fixed height (30 DMD pixels) and width (48 DMD pixels). On the left and right shows the initial, the 100[th] iteration and the optimized mask displayed on the DMD.

**Combining front-end programming and back-end sampling programming**
We then performed the joint programming of the front-end (wavefront shaping of the input data that manipulates the nonlinear interaction of MMF [26]) and the single-shot mask optimization of the DMD at the MMF output. We obtained 89.9% classification accuracy, up from 86.9% with only the back-end single-shot DMD programming. This accuracy is remarkably the same as a pretrained EfficientNet-B6 architecture with 43322513 parameters and followed by a digital decision layer [30]. However, in our case, only 1003 training parameters were used, a reduction of 99% in training parameters. We summarize in the Table below the performance achieved by the different approaches, digital network and optical network, in terms of the classification accuracy and the number of training parameters.

**Table 1.** Summary of performance of optical and digital implementations of networks for the classification of the COVID-19 x-ray images database. For the optical system, the color code indicates the type of programmable parameters: Blue = digital decision layer, Red = DMD superpixels, Green = front-end programming.

| Configuration | # of Training Parameter | Test Accuracy |
|---|---|---|
| SOLO [24] | (22×22)+1 | 83% |
| 4 horizontally sweeping multi-line DMD (4H) | 4×(22×22)+1 =1,937 | 86.2% |
| 5 vertically sweeping multi-line DMD (5V) | 5×(22×22)+1 =2,421 | 85.9% |
| LeNet-5 | 82,826 | 87.3% |
| 4H+5V | 9×(22×22)+1 =4,357 | 88.0% |
| 4H + optimized distance | 4×(22×22)+1 =1,937 | 88.0% |
| Single-shot DMD | (30×30+1)+80 =981 | 86.9% |
| EfficientNet-B6 | 43,322,513 | 89.9% |
| Single shot DMD + front-end programming [26] | (30×30+1)+52+50 =1,003 | 89.9% |

**Conclusion**

The P-SOLO system in this work was achieved by spatial-spectral optimization based on DMD sampling and grating dispersion placed at the MMF output. We optimized the binary mask on the DMD to extract important features for classification tasks. The advantage of our P-SOLO system was compared and demonstrated with respect to digital neural networks. The performance of the P-SOLO system might be further improved with mechanically perturbed MMF through a polarization control. By combining the wavefront shaping method in [26] and our proposed sampling method at the MMF output, we achieved a remarkable classification accuracy of 89.9% while reducing training parameters by 99% compared to an equally performing digital neural network.

In summary, we have experimentally demonstrated a P-SOLO system that can be used for a range of learning tasks classification. The programming single-shot spatial mask or multiple-line sweep method combined with spectral dispersion have achieved accuracy improvement for classifying COVID-19 X-ray images. The DMD mask only achieves amplitude modulation for the output beam. We anticipate that using amplitude and phase modulation simultaneously might further improve the performance.

**CRediT authorship contribution statement**

Y.Z. and J-L.H conceived the project. Y.Z, I.O. and J-L.H. built the system and performed the measurements. M.Y., N.U.D, C.G provided crucial suggestions for the main concept of the project. K. K. Y. W., C.M. and D.P. supervised the findings of this work.

**Declaration of competing interest**
The authors declare no competing financial interest.

**Data availability**
The data that support the plots within this paper and other finding of this study are available from the corresponding author upon reasonable request.

**Acknowledgments**
This work was supported by the Google--EPFL Collaboration Grant (No.901381) and the Swiss National Science Foundation (SNSF 200021_188419).

**References**
[1] C. Zhang, P. Li, G. Sun, Y. Guan, B. Xiao, J. Cong, Optimizing FPGA-based Accelerator Design for Deep Convolutional Neural Networks, in: Proceedings of the 2015 ACM/SIGDA International Symposium on Field-Programmable Gate Arrays, Association for Computing Machinery, New York, NY, USA, 2015: pp. 161–170. https://doi.org/10.1145/2684746.2689060.
[2] A. Krizhevsky, I. Sutskever, G.E. Hinton, ImageNet Classification with Deep Convolutional Neural Networks, in: Advances in Neural Information Processing Systems, Curran Associates, Inc., 2012. https://papers.nips.cc/paper/2012/hash/c399862d3b9d6b76c8436e924a68c45b-Abstract.html (accessed February 20, 2023).
[3] Y. LeCun, Y. Bengio, G. Hinton, Deep learning, Nature 521 (2015) 436–444. https://doi.org/10.1038/nature14539.
[4] K. He, X. Zhang, S. Ren, J. Sun, Deep Residual Learning for Image Recognition, in: 2016: pp. 770–778. https://openaccess.thecvf.com/content_cvpr_2016/html/He_Deep_Residual_Learning_CVPR_2016_paper.html (accessed February 20, 2023).
[5] B. Marr, B. Degnan, P. Hasler, D. Anderson, Scaling Energy Per Operation via an Asynchronous Pipeline, IEEE Transactions on Very Large Scale Integration (VLSI) Systems 21 (2013) 147–151. https://doi.org/10.1109/TVLSI.2011.2178126.
[6] J.M. Shainline, S.M. Buckley, R.P. Mirin, S.W. Nam, Superconducting Optoelectronic Circuits for Neuromorphic Computing, Phys. Rev. Appl. 7 (2017) 034013. https://doi.org/10.1103/PhysRevApplied.7.034013.
[7] P.R. Prucnal, B.J. Shastri, eds., Neuromorphic Photonics, CRC Press, Boca Raton, 2017. https://doi.org/10.1201/9781315370590.
[8] D. Woods, T.J. Naughton, Photonic neural networks, Nature Phys 8 (2012) 257–259. https://doi.org/10.1038/nphys2283.
[9] D.R. Solli, B. Jalali, Analog optical computing, Nature Photon 9 (2015) 704–706. https://doi.org/10.1038/nphoton.2015.208.
[10] X. Lin, Y. Rivenson, N.T. Yardimci, M. Veli, Y. Luo, M. Jarrahi, A. Ozcan, All-optical machine learning using diffractive deep neural networks, Science 361 (2018) 1004–1008. https://doi.org/10.1126/science.aat8084.

[11] T. Yan, J. Wu, T. Zhou, H. Xie, F. Xu, J. Fan, L. Fang, X. Lin, Q. Dai, Fourier-space Diffractive Deep Neural Network, Phys. Rev. Lett. 123 (2019) 023901. https://doi.org/10.1103/PhysRevLett.123.023901.

[12] G.V. der Sande, D. Brunner, M.C. Soriano, Advances in photonic reservoir computing, Nanophotonics 6 (2017) 561–576. https://doi.org/10.1515/nanoph-2016-0132.

[13] L. Larger, A. Baylón-Fuentes, R. Martinenghi, V.S. Udaltsov, Y.K. Chembo, M. Jacquot, High-Speed Photonic Reservoir Computing Using a Time-Delay-Based Architecture: Million Words per Second Classification, Phys. Rev. X 7 (2017) 011015. https://doi.org/10.1103/PhysRevX.7.011015.

[14] J. Feldmann, N. Youngblood, C.D. Wright, H. Bhaskaran, W.H.P. Pernice, All-optical spiking neurosynaptic networks with self-learning capabilities, Nature 569 (2019) 208–214. https://doi.org/10.1038/s41586-019-1157-8.

[15] R. Hamerly, L. Bernstein, A. Sludds, M. Soljačić, D. Englund, Large-Scale Optical Neural Networks Based on Photoelectric Multiplication, Phys. Rev. X 9 (2019) 021032. https://doi.org/10.1103/PhysRevX.9.021032.

[16] M. Miscuglio, Z. Hu, S. Li, J.K. George, R. Capanna, H. Dalir, P.M. Bardet, P. Gupta, V.J. Sorger, Massively parallel amplitude-only Fourier neural network, Optica, OPTICA 7 (2020) 1812–1819. https://doi.org/10.1364/OPTICA.408659.

[17] J. Bueno, S. Maktoobi, L. Froehly, I. Fischer, M. Jacquot, L. Larger, D. Brunner, Reinforcement learning in a large-scale photonic recurrent neural network, Optica, OPTICA 5 (2018) 756–760. https://doi.org/10.1364/OPTICA.5.000756.

[18] X. Xu, M. Tan, B. Corcoran, J. Wu, A. Boes, T.G. Nguyen, S.T. Chu, B.E. Little, D.G. Hicks, R. Morandotti, A. Mitchell, D.J. Moss, 11 TOPS photonic convolutional accelerator for optical neural networks, Nature 589 (2021) 44–51. https://doi.org/10.1038/s41586-020-03063-0.

[19] T. Wang, S.-Y. Ma, L.G. Wright, T. Onodera, B.C. Richard, P.L. McMahon, An optical neural network using less than 1 photon per multiplication, Nat Commun 13 (2022) 123. https://doi.org/10.1038/s41467-021-27774-8.

[20] D. Psaltis, D. Brady, K. Wagner, Adaptive optical networks using photorefractive crystals, Appl. Opt., AO 27 (1988) 1752–1759. https://doi.org/10.1364/AO.27.001752.

[21] L.G. Wright, T. Onodera, M.M. Stein, T. Wang, D.T. Schachter, Z. Hu, P.L. McMahon, Deep physical neural networks trained with backpropagation, Nature 601 (2022) 549–555. https://doi.org/10.1038/s41586-021-04223-6.

[22] T. Zhou, X. Lin, J. Wu, Y. Chen, H. Xie, Y. Li, J. Fan, H. Wu, L. Fang, Q. Dai, Large-scale neuromorphic optoelectronic computing with a reconfigurable diffractive processing unit, Nat. Photonics 15 (2021) 367–373. https://doi.org/10.1038/s41566-021-00796-w.

[23] J. Spall, X. Guo, A.I. Lvovsky, Hybrid training of optical neural networks, Optica, OPTICA 9 (2022) 803–811. https://doi.org/10.1364/OPTICA.456108.

[24] U. Teğin, M. Yıldırım, İ. Oğuz, C. Moser, D. Psaltis, Scalable optical learning operator, Nat Comput Sci 1 (2021) 542–549. https://doi.org/10.1038/s43588-021-00112-0.

[25] G.-B. Huang, Q.-Y. Zhu, C.-K. Siew, Extreme learning machine: Theory and applications, Neurocomputing 70 (2006) 489–501. https://doi.org/10.1016/j.neucom.2005.12.126.

[26] I. Oguz, J.-L. Hsieh, N.U. Dinc, U. Teğin, M. Yildirim, C. Gigli, C. Moser, D. Psaltis, Programming nonlinear propagation for efficient optical learning machines, AP 6 (2024) 016002. https://doi.org/10.1117/1.AP.6.1.016002.

[27] D. Eriksson, D. Bindel, C.A. Shoemaker, pySOT and POAP: An event-driven asynchronous framework for surrogate optimization, (2019). https://doi.org/10.48550/arXiv.1908.00420.

[28] COVID-19 Radiography Database, (n.d.).
https://www.kaggle.com/datasets/tawsifurrahman/covid19-radiography-database
(accessed March 4, 2024).
[29] Y. Lecun, L. Bottou, Y. Bengio, P. Haffner, Gradient-based learning applied to document recognition, Proceedings of the IEEE 86 (1998) 2278–2324. https://doi.org/10.1109/5.726791.
[30] M. Tan, Q. Le, EfficientNet: Rethinking Model Scaling for Convolutional Neural Networks, in: Proceedings of the 36th International Conference on Machine Learning, PMLR, 2019: pp. 6105–6114. https://proceedings.mlr.press/v97/tan19a.html (accessed February 20, 2023).

**Supplementary**
**Line width of DMD line mask:**
For the COVID-19 dataset of 3000 X-ray samples, the accuracy evolution for different horizontal linewidth is shown in Fig. S1(a). The best accuracy is 86.2% for a linewidth of 48 DMD pixels, corresponding to 4 intensity images in the horizontal direction. Similarly, the vertical direction optimized linewidth is 58 DMD pixels corresponding to a classification accuracy of 85.9% (Fig. S1(b)). Five different intensity images correspond to a vertical five-line sequentially swept at different locations of the MMF output beam.

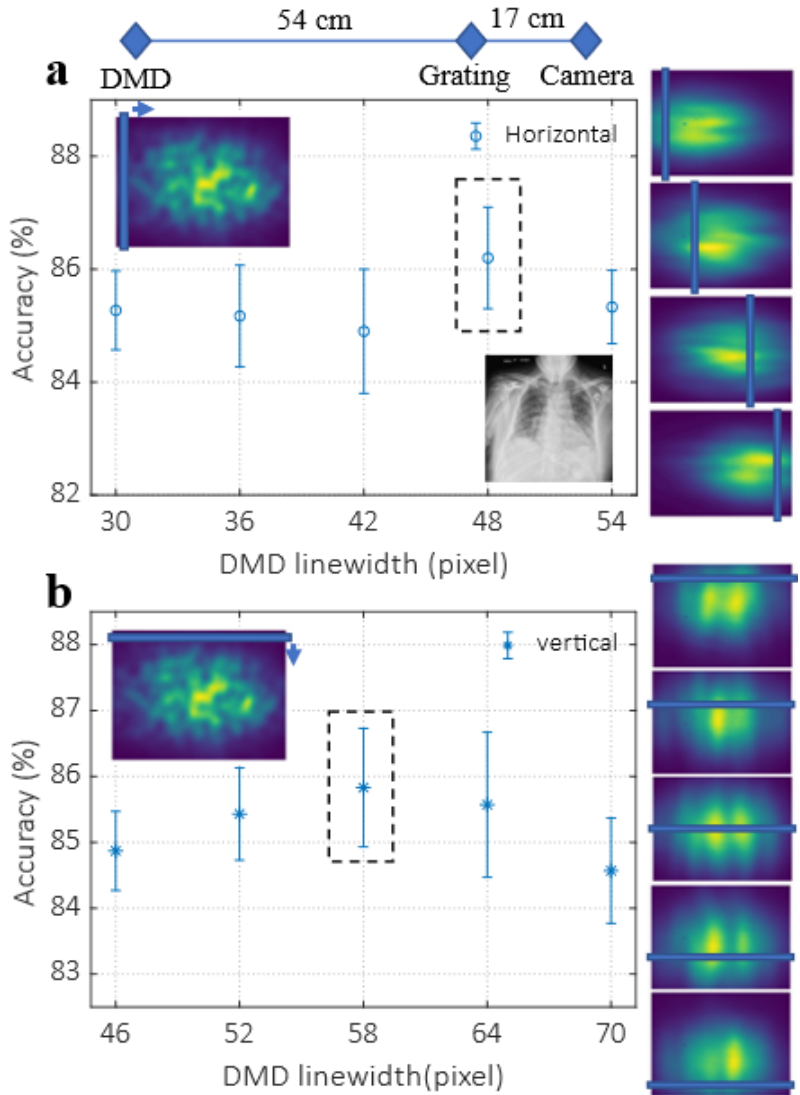

**Fig. S1.** Classification accuracy of COVID-19 X-ray dataset (3000 images) with optimization DMD mask line width. Accuracy variation of different DMD mask linewidth in (a) horizontal direction. (b) vertical direction.

**Surrogate optimization algorithm:** The dependency between DMD regions and classification accuracy is modeled by a surrogate model algorithm. By tuning the patterns around possible optimum, the algorithm creates the model of the system and finds the global optimum DMD sampling mask. For our implementation, we use Python Surrogate Optimization Toolbox (pySOT) [27]. The objective function in the optimization is classification accuracy. The variables that need to be optimized are the on/off state of DMD pixels. During the surrogate optimization, an initial population of 80 microregions is created corresponding to the single-shot mask on the DMD. The objective value of specific classification accuracy is measured for each DMD mask pattern during each iteration. Subsequently, the program iteratively optimizes the DMD mask by sampling parameter sets with higher expected improvement.